\documentclass[conference]{IEEEtran}

\usepackage{amssymb}
\usepackage{amsmath}
\usepackage{graphicx}
\usepackage{cite}
\usepackage{subfigure}
\usepackage{graphicx,epstopdf}
\usepackage{epsfig}	
\usepackage{cite,graphicx,amsmath,amssymb,amsthm}
\usepackage{comment}
\usepackage{nicefrac} % to make the fraction smaller \sfrac
\usepackage{lipsum}

\usepackage{bm}
\usepackage{textcomp}

\usepackage{algorithm}
\usepackage[noend]{algpseudocode}
\usepackage{indentfirst}

\makeatletter
\def\BState{\State\hskip-\ALG@thistlm}
\makeatother

\usepackage{amssymb}
\usepackage{amsmath}
\usepackage{graphicx}
\usepackage{cite}
\usepackage{citesort}
\IEEEoverridecommandlockouts

\usepackage{graphicx,epstopdf}
\usepackage{epsfig}	
\usepackage{amsfonts,balance}
\usepackage{bbm}
\floatname{algorithm}{Algorithm}
\usepackage{lipsum}

\newtheorem{theorem}{Theorem}

\newtheorem{corollary}{Corollary}

\makeatletter
\def\ScaleIfNeeded{%
\ifdim\Gin@nat@width>\linewidth \linewidth \else \Gin@nat@width
\fi } \makeatother

\begin{document}

\title{Weighted Sum-Rate Maximization for Rate-Splitting Multiple
Access Based Secure Communication}
\author{
\IEEEauthorblockN{ Huiyun Xia\IEEEauthorrefmark{1}, Yijie Mao\IEEEauthorrefmark{2}, Bruno Clerckx\IEEEauthorrefmark{3}, Xiaokang Zhou\IEEEauthorrefmark{1}, Shuai Han\IEEEauthorrefmark{1}
and Cheng Li\IEEEauthorrefmark{4}
% and Karen C. Cheung\IEEEauthorrefmark{2}  
%and
%Robert Schober\IEEEauthorrefmark{2}\IEEEauthorrefmark{4}
 } \IEEEauthorblockA{
\IEEEauthorrefmark{1}School of Electronics and Information Engineering, Harbin Institute of Technology, Harbin, China\\
\IEEEauthorrefmark{2}School of Information Science and Technology, ShanghaiTech University, Shanghai, China \\
\IEEEauthorrefmark{3}Department of Electrical and Electronic Engineering, Imperial College London, London, UK\\
\IEEEauthorrefmark{4}Electrical and Computer Engineering Faculty of Engineering and Applied Science, Memorial University, St. John's, Canada\\
 }
% \vspace{-0.6cm}
% \thanks{This work was supported by the Natural Science Foundation of China under Grant 61831002. The work of Huiyun Xia was supported by the China Scholarship Council. }
  }

\maketitle

\begin{abstract}
As investigations on physical layer security evolve from point-to-point systems to multi-user scenarios, multi-user interference (MUI) is introduced and becomes an unavoidable issue. Different from treating MUI totally as noise in conventional secure communications, in this paper, we propose a rate-splitting multiple access (RSMA)-based secure beamforming design, where user messages are split and encoded into common and private streams. Each user not only decodes the common stream and the intended private stream, but also tries to eavesdrop the private streams of other users. We formulate a weighted sum-rate (WSR) maximization problem subject to the secrecy rate requirements of all users. To tackle the non-convexity of the formulated problem, a successive convex approximation (SCA)-based approach is adopted to convert the original non-convex and intractable problem into a low-complexity suboptimal iterative algorithm. Numerical results demonstrate that the proposed secure beamforming scheme outperforms the conventional multi-user linear precoding (MULP) technique in terms of the WSR performance while ensuring user secrecy rate requirements.
\end{abstract}


\maketitle

\section{Introduction}

% introduce improtance of security 
Wireless users are facing unprecedented information security threats due to the broadcast nature of wireless channels. Consequently, information confidentiality has drawn critical research concerns recently \cite{Lv20215Gsecurity,Sai2021IRS}. Over decades, computation-demanding encryption algorithms have been the prevailing solution. However, to combat adversarial users with ever-increasing computation capability, complicated secret keys result in an unavoidable overconsumption of energy and computation resources \cite{xie2021PLAsurvey}. Meanwhile, those algorithms, requiring intricate key management and key distribution, may lead to poor performance in delay and outage probability \cite{rice2020V2EPLS}. By contrast, benefiting from the intrinsic randomness of wireless channels and noise, physical layer security becomes a hot spot field of research and serves as an efficient supplementary solution to assist data encryption thanks to its merits in theoretically information-secure performance and low complexity. 

% introduce physical layer security, its development, current hot-spot topic
As physical layer security evolves from the original point-to-point systems to multi-user systems, interference is inevitable when different transmit data share time-frequency resources to improve communication efficiency. Therefore, many researchers turn to study the effect of multiuser interference (MUI) on secure communication. Contrary to the opinion that treat MUI as a bottleneck of system performance, MUI generates two contrasting effects on communication security: on the one hand, it impedes a certain legitimate user from receiving its desired data due to malicious interception from other signals; on the other hand, MUI also adds difficulties for eavesdroppers to wiretap confidential data. Consequently, properly managed MUI can be leveraged to enhance communication security. Considering this, many works have been devoted to explore how to benefit secure communications from MUI in intercepting undesired users and in the meantime avoid the performance loss of desired users. For instance, researchers in \cite{cao2018secureprimarytransmission} and \cite{guo2019IAjamming} proposed interference alignment (IA) to eliminate the interference to undesired legitimate users and meanwhile disrupt eavesdroppers by jointly designing beamforming matrices of transceivers. 
%Then, with the assistance of artificial noise (AN) and jamming, the power allocation and beamforming design were analyzed in \cite{Xia2021IA} and \cite{guo2019IAjamming}, respectively. 
%Furthermore, its application was extended to Massive MIMO networks \cite{zhang2018IAmassivemimo}, Hyper-Dense Networks \cite{zhao2018hyperdense} and relay networks \cite{deeb2018relay}.
% On the other hand, authors in \cite{khandaker2018CI} proposed a constructive interference (CI) based secure precoding design, which takes advantage of interference and AN by reconstruct them from symbol level. Based on this, MUI and AN are jointly designed as constructive to the information receivers yet kept disruptive to potential eavesdroppers in securing simultaneous wireless information and power transfer (SWIPT) systems in \cite{khandaker2019constructiveinterference}. Later on, its application in distributed antenna systems as well as hardware implementation was discussed in \cite{wei2020secureprecoding} and \cite{wei2020interferencerole}, respectively. 

% introdce RS and its benefits on PLS
However, most current works adopt schemes where the designated confidential data stream is transmitted to its intended user, while the rest of legitimate users treat it as noise. Such methods can efficiently mitigate MUI, but may suffer from a performance loss in spectral efficiency (SE) and energy efficiency (EE). Regarding this, rate-splitting (RS) offers a new potential in managing MUI in multi-user multi-antenna communication networks by treating part of the interference as common streams to be decoded and the rest as noise, which is different from conventional interference management techniques \cite{Bruno2016RS}.
% add a paragraph on RS and RSMA in multi-user multi-antenna networks.
RSMA enables user messages to be split into certain common and private messages, after which the encoded common stream is superposed on top of all users’ private streams. The superposed streams are then precoded and transmitted from the multi-antenna transmitter. At the receiver side, successive interference cancellation (SIC) is applied at each user to enable sequential decoding of the common stream and the intended private stream. Receivers then reconstruct the original messages by their extracted part of the decoded common message and their separate private message \cite{Bruno2016RS}. RSMA therefore enables a more flexible interference management approach of partially decoding the interference and partially treating the interference as noise, and it has proved to be capable of bridging and outperforming space division multiple access (SDMA) and power-domain non-orthogonal multiple access (NOMA) \cite{mao2018rsma,bruno2020rs,bruno2021rs,mao2020rs,Mao2022rssurvey}. Furthermore, the performance advantage of RSMA has also been demonstrated in many aspects, such as robustness \cite{Hamdi2016robustRS}, SE and EE \cite{mao2019RSSEEE}. 

% literature review on RS+PLS
The investigation on RSMA-based physical layer security is still in its early stage. Recently,  \cite{hao2020robustsecureRS} studied RSMA-based two-user multiple input single output (MISO) secure communications by treating the common stream as both artificial noise (AN) to confuse the eavesdroppers and useful data for legal users to improve sum-rate. A robust secure resource allocation strategy was then proposed for user fairness optimization. Similarly, a cooperative rate-splitting (CRS) technique was investigated in a three-node secure RSMA-based relay channel in \cite{Ping2020cooperativeRS}. Besides, \cite{lu2021secureSWIPT} proposed a robust beamforming design to explore the worst-case EE performance. Under the aid of AN, a RSMA-based secure design was investigated in \cite{cai2021secureRA} to maximize the secrecy sum-rate by jointly optimizing beamforming and power allocation. 
Overall, RSMA is expected to enable a better management of MUI for intercepting potential adversarial user and enhancing legal users' performance. However, to the best of our knowledge, although there exist studies proving RSMA-based physical layer security schemes outperform traditional ones, most current studies focus on anti-eavesdropping scenarios, where
the eavesdroppers are not the intended recipients of messages from the transmitter, but only intercept confidential messages sent to other authorized users.
It is still not clear how the system performance may suffer when there exist internal eavesdroppers both receiving its intended message and meanwhile wiretapping messages intended for other users.

In this work, we bridge the gap by taking all user's secrecy rate constraints into consideration and we aim at optimizing beamforming design to maximize weighted sum-rate (WSR) in RSMA-based secure communications. We first construct a RSMA-based MISO broadcast channel (BC) for secure communication. By splitting user data into common streams for all users and private streams for separate users, we formulate a WSR maximization problem subject to the secrecy rate constraints at all users. To handle the non-convexity of the formulated optimization problem, a successive convex approximation (SCA)-based approach is adopted to obtain the optimal beamforming design. Finally, we compare the WSR performance by numerical results, which illustrate the performance advantage of our proposed beamforming design over conventional MULP design. 
% contributions

% The rest of this paper is organized as follows. Section II introduces the system model. The formulated secure beamformer design for WSR maximization problem and the proposed SCA approach is then specified in Section III. Numerical results are illustrated in Section IV. Finally, Section V concludes this paper.

% \emph{Notation:} Boldfaced lowercase letters and uppercase letters denote column vectors and matrices, respectively; ${\mathbb{C}}$ denotes the complex set; $(\cdot)^{\text{T}}$ and $(\cdot)^{\text{H}}$ denote the transpose and conjugate transpose operation, respectively; And $\text{Tr}\{\mathbf{A}\}$ denotes the trace of $\mathbf{A}$. Finally, $\mathbb{E\{\cdot\}}$ denotes the mathematical expectation.

\section{System Model}
\begin{figure*}[t]
\centerline{\includegraphics[scale=0.7]{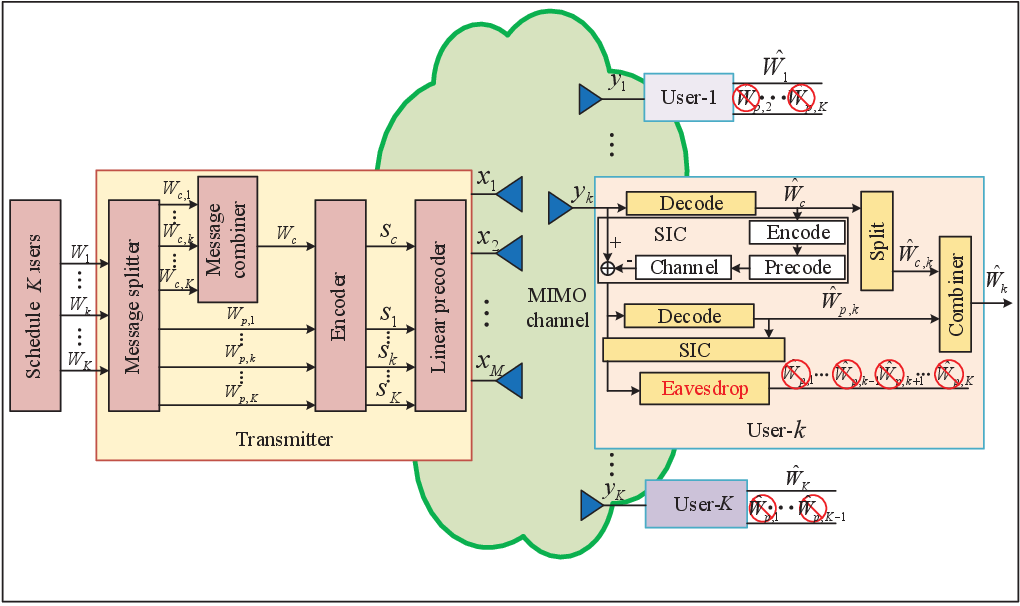}}
\caption{The system model of 1-layer RS based secure MISO BC.}
\label{secure_RSMA_MISO_BC}
\end{figure*}

As per Fig. \ref{secure_RSMA_MISO_BC}, consider a MISO BC secure communication model based on 1-layer RS, referred to as RS in the following, where one base station (BS) equipped with $N_t$ transmit antennas simultaneously serves $K$ single-antenna users, indexed by $\mathcal{K} = \{1,...,K\}$. The BS transmits $K$ confidential messages $W_1,...,W_K$ to the $K$ users. Each message $W_k$ is intended for the $k$th user, denoted by user-$k$, and needs to be kept secret from other users. Each user acts not only as a legitimate user to decode its desired message but also as a potential eavesdropper to wiretap messages transmitted to other users.

Following the principle of RS,  the message $W_k$ intended for user-$k$ is split into a common part $W_{{\rm c},k}$ and a private part $W_{{\rm p},k}$. The common parts of all users $W_{{\rm c},1}, . . . , W_{{\rm c},K}$ are then combined into a common message $W_{\rm c}$, which is encoded into the common stream $s_{\rm c}$ using a codebook shared by all users. The common stream $s_{\rm c}$ should be decoded by all users as it contains part of the messages of all users. Meanwhile, the private part $W_{{\rm p},k}$ of user-$k$ is independently encoded into a private stream $s_k$, which is only required to be decoded by user-$k$. The encoded common stream and private streams are then linearly precoded and the resulting transmit signal is
\vspace{-0.1cm}
\begin{equation}\label{1}
\mathbf{x}=\mathbf{P}\mathbf{s}=\mathbf{p}_{\rm c} s_{\rm c}+\sum_{k\in\mathcal{K}}\mathbf{p}_k s_k,
\end{equation}
where ${\bf{s}} = [s_{\rm c}, s_1, s_2, . . . , s_K]^{\rm{T}}$ is the grouped symbols in a given channel use. We assume $\mathbb{E}[\bf{s}\bf{s}^{\rm{H}}] = \bf{I}$. Therefore, the transmit power constraint is given as ${\rm{tr}}({\bf P}{\bf P}^{\rm{H}}) \le P_t$, where ${\bf P} = [{\bf p}_{\rm c}, {\bf p}_1, {\bf p}_2, . . . , {\bf p}_K]$ is the integrated precoder matrix and ${\bf p}_{\rm c}, {\bf p}_k \in {\mathbb C}^{N_t\times1}$ is the precoder for the common stream $s_{\rm c}$ and the $k$th private stream $s_k$, respectively. 
The received signal at user-$k$ for a given channel state is ${y}_k = \mathbf{h}^\text{H}_k \mathbf{x} + n_k$,
where $\mathbf{h}_k \in\mathbb{C}^{N_t\times1}$ is the channel between BS and user-$k$. We assume perfect CSI is available at the receivers and the transmitter. 
$n_k\sim\mathcal{CN}(0, \sigma_{n,k}^2)$ is the additive white Gaussian noise (AWGN) at user-$k$. Without loss of generality, we assume equal noise variances for all users, i.e. $\sigma_{n,k}^2 = \sigma_{n}^2, \forall k \in \cal{K}$. 

At the receiver side, each user-$k$ first decodes the common stream $s_{\rm c}$ into $\widehat W_{{\rm c}}$ by treating the interference from all the private streams as noise and then extracts its own part of common message $\widehat W_{{\rm c},k}$ from $\widehat W_{{\rm c}}$. With the assistance of ideal SIC, the decoded common message $\widehat W_{{\rm c},k}$ is then re-encoded, precoded and removed from the received signal. After that, each user decodes the intended private stream $s_k$ into $\widehat W_{{\rm p},k}$ by treating the interference from all other private streams as noise. User-$k$ then recovers the original message by combining $\widehat W_{{\rm c},k}$ with $\widehat W_{{\rm p},k}$ into $\widehat W_k$.
Consequently, at user-$k$, the corresponding achievable rates of common stream $R_{{\rm c},k}$ and private stream $R_{{\rm p},k}$ per channel use are given by
\vspace{-0.1cm}
\begin{subequations}
\begin{align}\label{3}
R_{{\rm c},k} &= \text{log}_2{(1 + \Gamma_{{\rm c},k} )}, \\ 
R_{{\rm p},k} &= \text{log}_2{(1 + \Gamma_{{\rm p},k} )}, 
\end{align}
\end{subequations}
where $\Gamma_{{\rm c},k}$ and $\Gamma_{{\rm p},k}$ are the respective signal to interference plus noise ratio (SINR) of the common stream and the private stream at user-$k$. They can be expressed as
\vspace{-0.1cm}
\begin{subequations}
\begin{align}
\Gamma_{{\rm c},k}&=\frac{| {{\bf{h}}_k^{\rm{H}}{{\bf{p}}_{\rm{c}}}} |^2}{\sum\limits_{i \in \mathcal{K}} {{{| {{\bf{h}}_k^{\rm{H}}{{\bf{p}}_i}} |}^2}}  + \sigma _n^2}, \label{4a}\\
\Gamma_{{\rm p},k}&=\frac{{| {{\bf{h}}_k^{\rm{H}}{{\bf{p}}_{k}}} |^2}}{\sum\limits_{i \in \mathcal{K}\backslash\{k\}} {{{| {{\bf{h}}_k^{\rm{H}}{{\bf{p}}_i}} |}^2}}  + \sigma _n^2}. \label{4b}
\end{align}
\end{subequations}

After decoding $s_k$ into $\widehat W_{{\rm p},k}$, one more layer of ideal SIC is employed to re-encode, precode, and remove the intended private stream from the received signal. User-$k$ then tries
to eavesdrop the private messages $\{W_{{\rm p},j}| j\in\mathcal{K}\backslash\{k\}\}$ intended for other users. Therefore, the achievable wiretap rate of decoding the private stream $s_j$ at user-$k$ per channel use is 
\vspace{-0.3cm}
\begin{equation}\label{5}
R_{j, k}=\log _{2}(1+\Gamma_{j,k}), \quad \forall j \in \mathcal{K} \backslash\{k\},
\end{equation}
where $\Gamma_{j,k}$ is the SINR of user-$j$'s private stream at user-$k$,
\vspace{-0.3cm}
\begin{equation}\label{gamma_jk}
\Gamma_{j, k}=\frac{|\mathbf{h}_{k}^{\rm H} \mathbf{p}_{j}|^{2}}{\sum\limits_{k' \in \mathcal{K}\backslash\{k,j\}}|\mathbf{h}_{k}^{\rm H} \mathbf{p}_{k'}|^{2}+ \sigma _n^2}, \quad \forall j \in \mathcal{K} \backslash\{k\}.
\end{equation}
The secrecy rate of user-$k$ is then defined as $R_{s,k}$=$[R_{{\rm P},k}-\max_{j \in \mathcal{K} \backslash\{k\}}\{R_{k,j}\}]^+$, where $\max_{j \in \mathcal{K} \backslash\{k\}}\{R_{k,j}\}$ is the largest wiretap rate at all potential eavesdroppers  to decode the private message for user-$k$. The operation $[x]^+=\max\{x,0\}$.

Notably, to guarantee $W_{\rm c}$ is successfully decoded by all users, the actual transmission rate $R_{\rm c}$ for the common stream should not exceed the minimum achievable rate of the common stream at all users, i.e. $\min_{k \in \mathcal{K}} R_{{\rm c},k}$. Moreover, $R_c$ is shared by all users in set $\mathcal{K}$ where each user-$k$ is allocated a portion $C_k$ corresponding to the rate of $W_{{\rm c},k}$. The common stream rate allocation $C_1,...,C_K$ satisfies $\sum_{k\in\mathcal{K}}C_k=R_{\rm c}$. Hence, the total achievable rate of user-$k$ is defined as $R_{k,tot} = C_k+R_{{\rm p},k}$.

%%%%%%%%%%%%%%%%%%%%%Section III%%%%%%%%%%%%%%%%%%%%%%%
\section{Problem Formulation and Optimization Framework}
\subsection{Problem Formulation}
We consider the secure RS precoder optimization problem with the aim of maximizing WSR of all users subject to the secrecy rate constraint of each user. For a given user weight vector ${\mathbf{u}}=[u_1,....,u_K]$, the optimization problem is given as
\vspace{-0.3cm}
\begin{subequations}
\begin{align}
{\rm{P0:}}\max _{\mathbf{c}, \mathbf{P}} & \sum_{k \in \mathcal{K}} u_{k}(C_{k}+R_{{\rm p},k}) \label{7a}\\
\text { s.t. } & R_{s, k} \geq R_{s,k}^{th}, \forall k \in \mathcal{K}, {\label{7b}}\\
& \sum_{j \in \mathcal{K}} C_{j} \leq R_{c, k}, \forall k \in \mathcal{K}, \label{7c}\\
& \operatorname{tr}(\mathbf{P} \mathbf{P}^{H}) \leq P_{t}, \label{7d}\\
& \mathbf{c} \geq \mathbf{0}, \label{7e}
\end{align}
\end{subequations}
where ${\mathbf{c}} = [C_1,...,C_K]$ is the common rate allocation vector specifying the rate of the common stream allocated to each user. Constraint \eqref{7b} is the secrecy rate constraint of each user. $R_{s,k}^{\rm {th}}$ is the secrecy rate threshold of each private stream. Constraint \eqref{7c} ensures each user successfully decode the common stream. Constraint \eqref{7d} is the transmit power constraint.

\subsection{Proposed Algorithm}
The non-convexity of P0 arises from \eqref{7a}-\eqref{7c}. Therefore, to find feasible solutions to this problem, a SCA-based method is proposed in this subsection.

Introducing $\boldsymbol\alpha=\{\alpha_{{\rm c},k},\alpha_{{\rm p},k},\alpha_{k,j}|  \forall k\in\mathcal{K}, j\in\mathcal{K}\backslash\{k\}\}$, ${\boldsymbol\rho}=\{\rho_{{\rm c},k},\rho_{{\rm p},k},\rho_{k,j}| \forall k\in\mathcal{K},j\in\mathcal{K}\backslash\{k\}\}$, and plugging \eqref{3}, \eqref{5} into $\rm P0$, \eqref{7b}, \eqref{7c} become
\vspace{-0.3cm}
\begin{subequations}
\begin{align}
&[\alpha_{{\rm p},k}-\alpha_{k,j}]^+\ge R_{s,k}^{\rm {th}}, k\in\mathcal{K}, j\in\mathcal{K}\backslash\{k\},	\label{8a}\\
&\sum_{j \in \mathcal{K}} C_{j} \leq \alpha_{{\rm c}, k}, \forall k \in \mathcal{K},\label{8b}\\
&1 + {\rho_{i,k}} \ge {2^{{\alpha _{i,k}}}}, \forall k\in\mathcal{K}, i\in\{{\rm c},{\rm p}\},	\label{8c}\\
&1 + {\rho_{k,j}} \le {2^{{\alpha _{k,j}}}}, \forall k\in\mathcal{K},\forall j\in \mathcal{K}\backslash\{k\}, \label{8d}\\
&{\Gamma_k} \ge {\rho_{{\rm p},k}}, \forall k\in\mathcal{K}, \label{8e}\\
&{\Gamma_{k,j}} \le {\rho_{k,j}}, \forall k\in\mathcal{K},\forall j\in \mathcal{K}\backslash\{k\}.\label{8f}
\end{align}
\end{subequations}
Plugging \eqref{4a}, \eqref{4b} into \eqref{8e}, \eqref{8f}, respectively, $\rm{P0}$ becomes
\vspace{-0.3cm}
\begin{subequations}
\begin{align}
{\rm {P1:}}&\max _{\mathbf{c}, \mathbf{P}, {\boldsymbol\alpha}, {\boldsymbol\rho}}  \sum_{k \in \mathcal{K}} u_{k}(C_{k}+\alpha_{{\rm p}, k}) \label{9a}\\
&\hspace{0.3cm}{\text {s.t. } } \frac{{{{| {{\bf{h}}_k^H{{\bf{p}}_c}} |}^2}}}{{\sum\limits_{j \in \mathcal{K}} {{{| {{\bf{h}}_k^H{{\bf{p}}_j}} |}^2}}  + \sigma_n^2}} \ge {\rho _{{\rm c},k}},\forall k\in\mathcal{K},\label{9c}\\
& \hspace{0.4cm} \frac{{{{| {{\bf{h}}_k^H{{\bf{p}}_k}} |}^2}}}{{\sum\limits_{j \in \mathcal{K}\backslash\{k\}} {{{| {{\bf{h}}_k^H{{\bf{p}}_j}} |}^2}}  + \sigma_n^2}} \ge {\rho _{{\rm p},k}},\forall k\in\mathcal{K},\label{9d}\\
& \hspace{0.4cm} \frac{{{{| {{\bf{h}}_j^H{{\bf{p}}_k}} |}^2}}}{{\sum\limits_{k' \in \mathcal{K}\backslash\{k,j\} } {{{| {{\bf{h}}_j^H{{\bf{p}}_{k'}}} |}^2}}  + \sigma_n^2}} \le {\rho _{k,j}}, \forall j \in \mathcal{K}\backslash\{k\}, \label{9e}\\
&\hspace{0.4cm} \nonumber \eqref{7d},\eqref{7e}, \eqref{8a}-\eqref{8d}. 
\end{align}
\end{subequations}

However, $\rm P1$ is still non-convex due to constraints \eqref{8d} and \eqref{9c}-\eqref{9e}. To address this problem, we adopt the first-order Taylor expansion to linearly approximate constraint \eqref{8d} as
\vspace{-0.1cm}
\begin{equation}\label{10}
1+\rho_{k,j}\le 2^{\alpha_{k,j}^{[n]}}[1+\ln2(\alpha_{k,j}-\alpha_{k,j}^{[n]})],~ \forall j\in\mathcal{K}/\{k\},
\end{equation}
where $\alpha_{k,j}^{[n]}$ denotes the optimized $\alpha_{k,j}$, obtained from the $[n]$-th iteration. 
Additionally, to reformulate constraints \eqref{9c} and \eqref{9d}, we introduce another vector $\boldsymbol\beta=\{\beta_{{\rm c},k},\beta_{{\rm p},k}\mid  \forall k\in\mathcal{K}\}$ to relax the denominator of the inequalities, and we obtain
\vspace{-0.1cm}
\begin{subequations}
\begin{align}
&\frac{{{{| {{\bf{h}}_k^H{{\bf{p}}_i}} |}^2}}}{{{\beta _{i,k}}}} \ge {\rho _{i,k}},\forall k \in {\cal K}, i\in\{{\rm c},{\rm p}\},	\label{11a}\\
&\sum\limits_{j \in {\cal K}} {{{| {{\bf{h}}_k^H{{\bf{p}}_j}}|}^2}}  + \sigma_n^2 \le {\beta _{{\rm c},k}},	\label{11b}\\
&\sum\limits_{j \in \mathcal{K}\backslash\{k\}} {{{| {{\bf{h}}_k^H{{\bf{p}}_j}} |}^2}}  + \sigma_n^2 \le {\beta _{{\rm p},k}}.	\label{11c}
\end{align}
\end{subequations}
Similarly, the non-convex constraints \eqref{11a} and \eqref{9e} can be reconstructed into the convex ones by adopting the first-order Taylor approximations, which are given as
\vspace{-0.1cm}
\begin{equation}\label{12}
\begin{aligned}
\frac{2 \Re\{(\mathbf{p}_{i}^{[n]})^{H} \mathbf{h}_{k} \mathbf{h}_{k}^{H} \mathbf{p}_{i}\}}{\beta_{i,k}^{[n]}} 
-\frac{|\mathbf{h}_{k}^{H} \mathbf{p}_{i}^{[n]}|^{2} \beta_{i,k}}{(\beta_{i,k}^{[n]})^{2}} \geq \rho_{i,k},\\
\forall k \in \mathcal{K}, i\in\{{\rm c},{\rm p}\}.
\end{aligned}
\end{equation}
and
% \begin{equation}\label{13}
% {{{{\left| {{\bf{h}}_j^H{{\bf{p}}_k}} \right|}^2}}}\le \rho_{k,j},\forall k\in\mathcal{K}, j\in\mathcal{K}\backslash\{k\}.
% \end{equation}
\vspace{-0.1cm}
\begin{equation}\label{13}
 \begin{aligned}
 & \rho_{_{k,j}}^{[n]}\sum\limits_{k' \in {\cal K}\backslash\{k,j\}}  \bigg({{2\Re} \big( {{{( {{\bf{p}}_{_{k'}}^{[n]}} )}^H}{{\bf{h}}_j}{\bf{h}}_j^H{{\bf{p}}_{k'}}} \big)} - {{{| {{\bf{h}}_j^H{\bf{p}}_{_{k'}}^{[n]}} |}^2}} \bigg)\\
 &+\rho_{k,j}\big( {\sum\limits_{k' \in {\cal K}\backslash\{k,j\}} {{{| {{\bf{h}}_j^H{\bf{p}}_{_{k'}}^{[n]}} |}^2}}  + \sigma_n^2} \big) \ge {| {{\bf{h}}_j^H{{\bf{p}}_k}} |^2}.
 \end{aligned}
 \end{equation}

Based on \eqref{10}, \eqref{12} and \eqref{13}, at the optimal solution
$({\mathbf P}^{[n-1]},{\boldsymbol\alpha}^{[n-1]}, {\boldsymbol\beta}^{[n-1]}, {\boldsymbol\rho}^{[n-1]})$ obtained from the $[n-1]$-th iteration, we obtain the following problem at the $[n]$-th iteration:
\vspace{-0.1cm}
\begin{subequations}
\begin{align}
{\rm{P2:}} & \max _{\mathbf{c}, \mathbf{P}, \boldsymbol{\alpha}, \boldsymbol{\beta}, \boldsymbol{\rho}}  \sum_{k \in \mathcal{K}} u_{k}(C_{k}+\alpha_{{\rm p}, k}) \label{14}\\
& \hspace{0.3cm} \text { s.t. } \hspace{0.2cm} \eqref{8a}-\eqref{8c}, \eqref{10}, \eqref{11b}, \eqref{11c}, \eqref{12}, \eqref{13},\eqref{7d},\eqref{7e}.\nonumber
\end{align}
\end{subequations}
$\rm P2$ is convex and can be effectively solved by applying state-of-art toolboxes, such as CVX. The proposed SCA-based algorithm is summarized as in Algorithm 1. At each iteration, the obtained parameters $({\mathbf P}^{[n-1]},{\boldsymbol\alpha}^{[n-1]}, {\boldsymbol\beta}^{[n-1]})$ from the last iteration serve as inputs for optimizing problem $\rm P2$ until the WSR converges to a preset threshold. 
\begin{algorithm}[!htbp]
\caption{SCA-based secure beamforming optimization}
\label{Algorithm 1}
\begin{algorithmic}[1]
    \Require
    the tolerance $\epsilon$, the secrecy threshold $R_{s,k}^{\rm th}$, the power constraint $P_t$.
    \Ensure
    the obtained WSR, the optimal secure precoder ${\mathbf P}^*$ and common rate vector ${\mathbf c}^*$ 
    \State Set threshold $\epsilon $, set iteration index $n:=0$;\\
     Initialize $({\mathbf P}^{[0]},{\boldsymbol\alpha}^{[0]}, {\boldsymbol\beta}^{[0]})$ and the WSR$^{[0]}$;
    \Repeat:
    \State \textit{Calculate.} Using $({\mathbf P}^{[n-1]},{\boldsymbol\alpha}^{[n-1]}, {\boldsymbol\beta}^{[n-1]})$ obtained \State from last iteration, solve problem $\rm P2$ and assign
    \State  the obtained optimized variables $({\mathbf P}^{\rm *},{\boldsymbol\alpha}^{\rm *}, {\boldsymbol\beta}^{\rm *})$ 
    \State as $({\mathbf P}^{[n]},{\boldsymbol\alpha}^{[n]}, {\boldsymbol\beta}^{[n]})$ and optimized objective value 
    \State WSR$^{\rm *}$ as WSR$^{[n]}$ 
    \State \textit{Update iteration.} $n=n+1$
    \Until \textit{Stopping criterion satisfying:} 

    $| {\rm WSR}^{[n]}-{\rm WSR}^{[n-1]} |\le \epsilon$   
    \State \Return ${\mathbf P}^*={\mathbf P}^{[n]}$, ${\mathbf c}^*={\mathbf c}^{[n]}$
\end{algorithmic}
\end{algorithm}

\emph{Initialization:}
We initialize the precoder ${\bf{P}}^{[0]}$ as in \cite{mao2018rsma}. The precoders for the private streams ${\bf p}_k$ are initialized by ${\bf p}_k=p_k\frac{{\bf h}_k}{\lVert{\bf h}_k\rVert}$, where $p_k=\frac{\lambda P_t}{K}$ and $0\leq\lambda\leq 1$. The precoder for the common streams is initialized by ${\bf p}_{\rm c}=p_{\rm c} {\bf u}_{\rm c}$, where  $p_{\rm c}=(1-\lambda)P_t$ and ${\bf u}_{\rm c}$ is the largest left singular vector of the channel matrix ${\bf H}=[ {\bf h}_1, ..., {\bf h}_K]$. The common rate vector ${\bf{c}}^{[0]}$ is initialized by assuming the rate of the common stream is uniformly allocated to all users. $\beta_{{\rm c},k}^{[0]}$, $\beta_{{\rm p},k}^{[0]}$ and $\alpha_{k,j}^{[0]}$ are initialized by replacing the inequalities of \eqref{11b}, \eqref{11c} and \eqref{8d} with equalities, respectively.    

 \emph{Convergence analysis:} 
As \eqref{10}, \eqref{12} and \eqref{13} are the lower bound approximations of \eqref{8d}, \eqref{11a} and \eqref{9e}, the optimal solution obtained in the $[n-1]$-th iteration also serves as a feasible solution for the $[n]$-th iteration. Therefore, the corresponding optimal objective value ${\rm WSR}^{[n-1]}$ is no larger than ${\rm WSR}^{[n]}$. Besides, the objective function is bounded by the transmit power constraint \eqref{7d}, the convergence is hence guaranteed. However, there is no guarantee of the global optimality since a local optimum can also result in termination of the iteration process.

\section{Numerical Results}
In this section, we evaluate the WSR performance of a 2-user secure RS communication and compare the proposed secure RS with MULP. Both settings when the number of transmit antennas is $N_t=2$ and $N_t=4$ are considered. We set the user weights ${\bf u}=[0.5, 0.5]$, and noise variance $\sigma_n^2=1$. We also assume all users have the same secrecy rate constraints. Notably, NOMA is not considered in our simulation. This is because the use of SIC at receivers forces some user data to be fully decoded at an unintended user, which fails to ensure the security of these user data. Besides, research has demonstrated NOMA is a subscheme of RSMA \cite{mao2018rsma,bruno2020rs}.
\subsubsection{Specific channels}
To investigate the influence of user angles and channel strength on security, following \cite{mao2018rsma}, a specific channel scenario is first analyzed. Taking $N_t=4$ as an example, the user channels are realized as
\vspace{-0.1cm}
\begin{equation}
\begin{aligned}
{\bf{h}}_1&=[1, 1, 1, 1]^H,\\
{\bf{h}}_2&=\gamma\times[1, e^{j\theta}, e^{j2\theta}, e^{j3\theta}]^H,
\end{aligned}
\end{equation}
where $\gamma$ and $\theta$ represent the relative channel strength and angle between user-2 and user-1, respectively. We set $\theta=[\frac{\pi}{9}, \frac{2\pi}{9}, \frac{3\pi}{9}, \frac{4\pi}{9}]$ and transmit power constraint $P_t=20$dB.  

%% (2,2)_specific_WSR,NT=2,K=2,BIAS=1,PT=20dB

 \begin{figure}[t]
 \centerline{\includegraphics[scale=0.5]{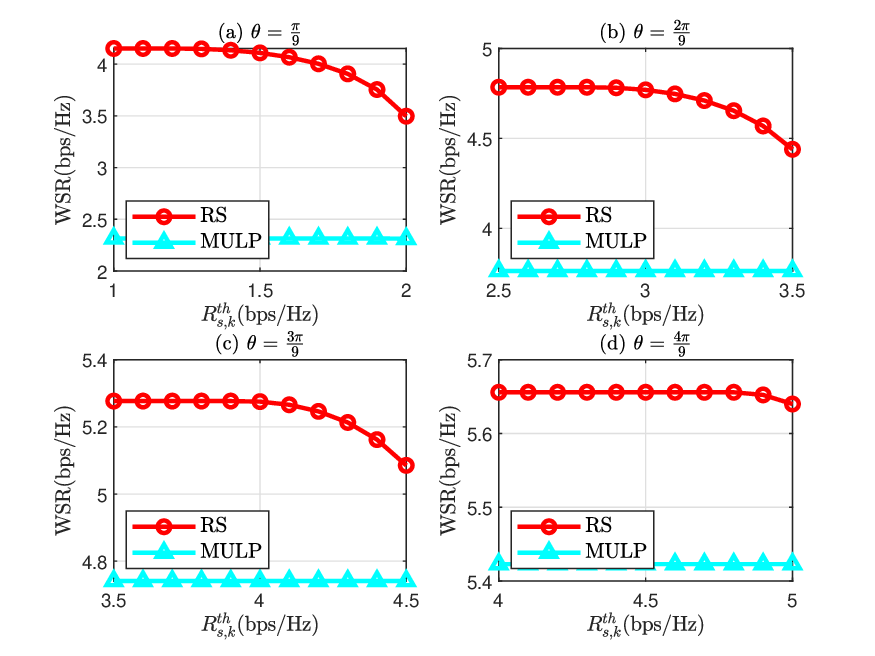}}
 \caption{The WSR vs $R_{s,k}^{th}$ for secure RS and MULP schemes, $N_t=2, \gamma=1$.} 
 \label{(2,2)_specific_WSR}
 \end{figure}

Fig. \ref{(2,2)_specific_WSR} illustrates the WSR performance of secure RS and traditional MULP schemes with respect to different secrecy threshold $R_{s,k}^{\rm th}$, when $N_t=2$, $\gamma=1$, that is, both user channels have the same channel strength. From this figure, the WSR of secure RS always outperforms that of MULP as the phase difference between user channels $\theta$ ranges from $\frac{\pi}{9}$ to $\frac{4\pi}{9}$, which means the user channels vary from almost aligned to almost orthogonal to each other, RS achieves better WSR than MULP. This is because RS is more flexible in optimizing precoders by bridging common rate and private rate. To be specific, the WSR of secure RS first remains unchanged then decreases, since when the secrecy rate constraint $R_{s,k}^{th}$ is relatively small, RS is capable of ensuring security by optimizing private streams without decreasing each user's WSR performance. However, as $R_{s,k}^{th}$ increases, more power should be allocated to private streams so as to enhance security though introducing WSR loss in the meantime. 
 Whereas the WSR of MULP still remains at a relatively low level compared with RS as no common streams are scheduled in this scheme. 

%% (2,2)_specific_PA,BIAS=1,PT=20dB
 \begin{figure}[t]
 \centerline{\includegraphics[scale=0.5]{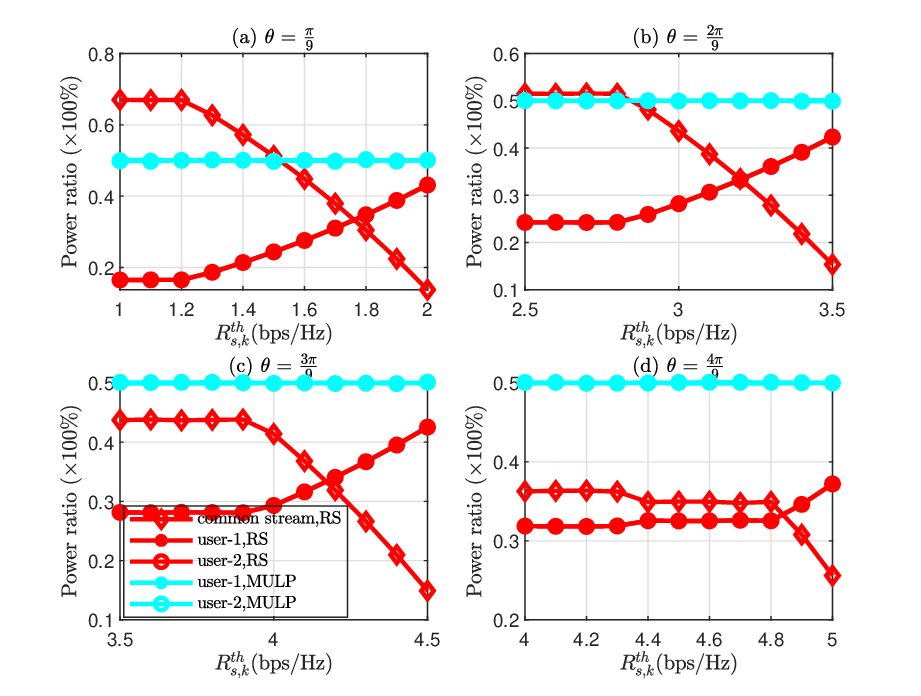}}
 \caption{The power allocation of data streams vs $R_{s,k}^{\rm th}$ for secure RS and MULP schemes, $N_t=2, \gamma=1$.}
 \label{(2,2)_specific_PA}
 \end{figure}
 Fig. \ref{(2,2)_specific_PA} illustrates the power ratio of common streams and private streams for secure RS and MULP scheme. We observe from the figure that the power ratio curves of private streams for user-1 and user-2 are overlapped for both MULP and RS, since $\gamma=1$ indicates both users have the same channel strength. Additionally, for secure RS scheme, as $R_{s,k}^{\rm th}$ increases, the power allocated to common stream first remains constant then decreases in all $\theta$ values. That is because a higher security constraint requires more power allocated to private streams, which corresponds to results in Fig. \ref{(2,2)_specific_WSR}. As there is no common stream in MULP, the power ratio for both private streams stay constant at 0.5.

 \begin{figure}[t]
 \centerline{\includegraphics[scale=0.5]{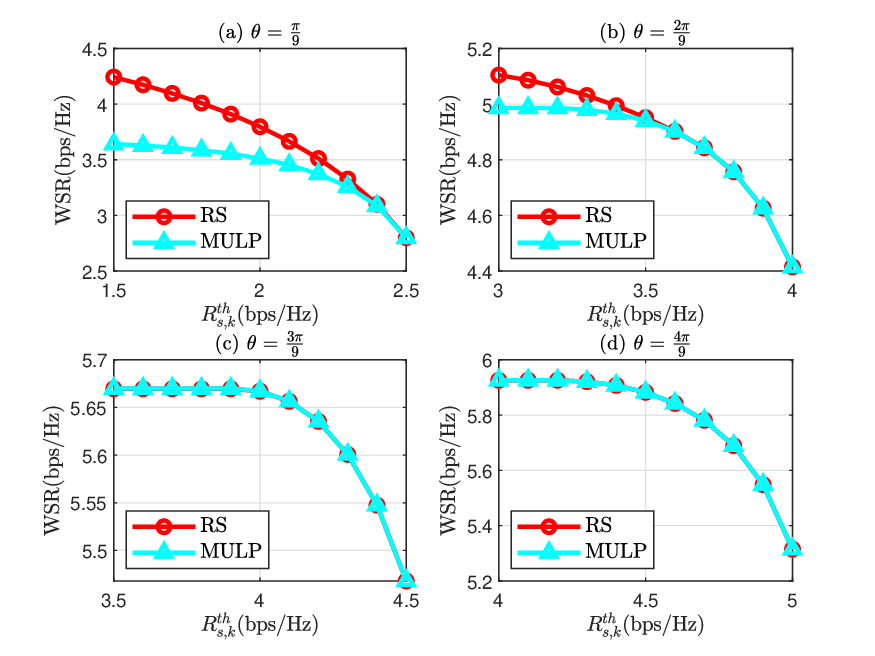}}
 \caption{The WSR vs $R_{s,k}^{\rm th}$ for secure RS and MULP schemes, $N_t=4, \gamma=0.3$.}
 \label{(4,2)_specific_b0.3_WSR}
 \end{figure}

 %% (4,2)_specific_PA,NT=4,K=2,BIAS=0.3, PT=20dB
\begin{figure}[t]
\centerline{\includegraphics[scale=0.5]{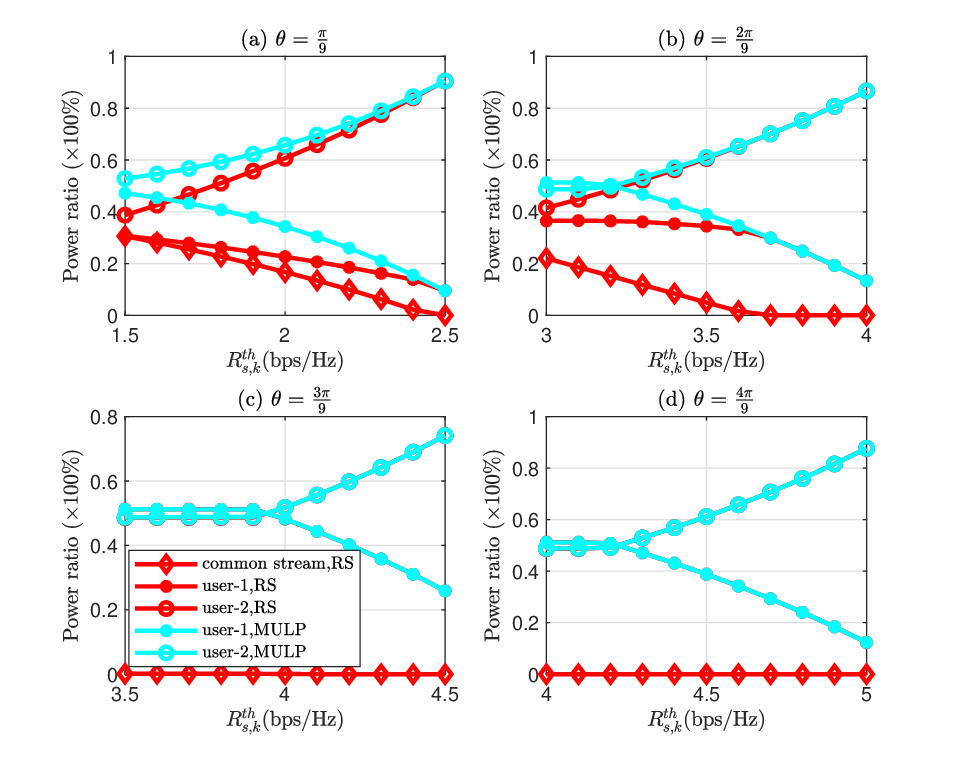}}
\caption{The power allocation of common streams and private streams vs $R_{s,k}^{\rm th}$ for secure RS and MULP schemes, when $N_t=4, \gamma=0.3$.}
\label{(4,2)_specific_b0.3_PA}
\end{figure}

 % \begin{figure}[htbp]
 % \centerline{\includegraphics[scale=0.6]{(4,2)_specific_WSR_Rth_bias=1.eps}}
 % \caption{The WSR versus secrecy threshold $R_{s,k}^{th}$ for 2-user secure RS and MULP schemes under specific channels, when $N_t=4, K=2, \gamma=1$, $P_t=20dB$.}
 % \label{fig3}
 % \end{figure}
 Fig. \ref{(4,2)_specific_b0.3_WSR} and Fig. \ref{(4,2)_specific_b0.3_PA} illustrate the WSR and power allocation of data streams with respect to different secrecy threshold $R_{s,k}^{\rm th}$ for both secure RS and MULP schemes, when $N_t=4$ and $\gamma=0.3$. That is, user-2 has 10 dB channel strength difference compared with user-1.
 We observe from Fig. \ref{(4,2)_specific_b0.3_WSR}, RS exhibits a performance advantage over MULP when user channels are relatively aligned and the security threshold is relatively small. Whereas when $\theta=\frac{3\pi}{9}$ and $\theta=\frac{4\pi}{9}$, these two schemes achieve the same WSR performance. This shows that as $\theta$ increases, channel orthogonality dominates in securing data streams. 
 % Correspondingly, as the security requirement increases, the power ratio of common streams becomes smaller and smaller, and finally turns to zero when channel orthogonality is ensured and that is also when the two WSR performance curves coincide.
As for the corresponding power allocation, it can be seen in Fig. \ref{(4,2)_specific_b0.3_PA}, the power allocated to common stream decreases as $R_{s,k}^{th}$ increases when $\theta=\frac{\pi}{9}$ and $\theta=\frac{2\pi}{9}$. This also explains why RS outperforms MULP in WSR performance in such case. Meanwhile, with fixed $\theta$, a higher security threshold tends to allocate more power to the private stream of the user with poorer channel strength. When user channels are approximately orthogonal to each other, both RS and MULP obtain the same power allocation results, therefore their corresponding WSR performance curves coincide.

% when the 2 user channels are relatively aligned to each other, the power ratio for user-2's private data is obviously larger than that of user-1. However, as $\theta$ increases, the difference become smaller and finally the power for private streams tend to be equally distributed between 2 users when $\theta=\frac{4}{9}$, which means in such condition, channel orthogonality can already ensure the required security without power allocation.
%%%%%%%%%%%%%%%%%%%%%%%%%%%%%
\subsubsection{Random Channels}
We compare the WSR performance of 2-user scenarios between secure RS and secure MULP schemes averaged by 100 random channels. Each element of the user channel follows the independent and identical zero-mean complex Gaussian distribution with unit variance, i.e., $\mathcal{CN}(0, 1)$.

 \begin{figure}[htbp]
 \centerline{\includegraphics[scale=0.45]{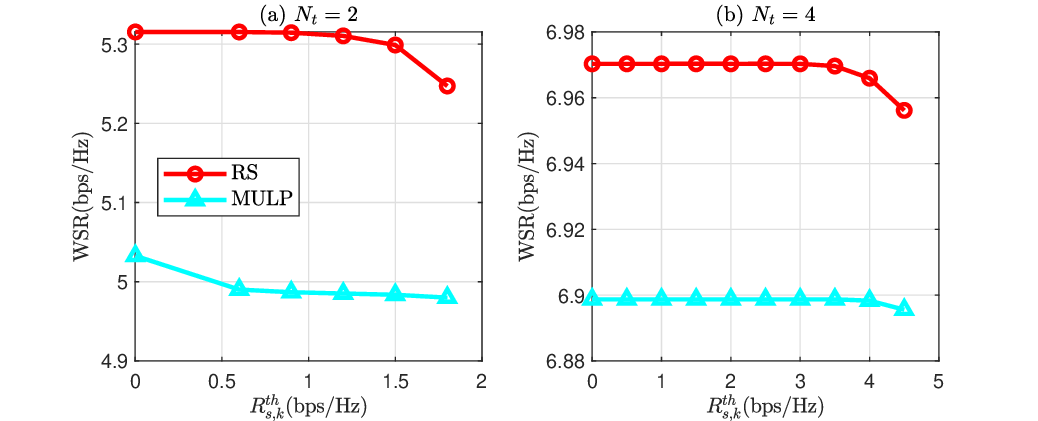}}
 \caption{The WSR versus secrecy threshold $R_{s,k}^{\rm th}$ over 100 random channels for 2-user secure RS and MULP, $P_t=20$dB, (a) $N_t=2$, (b) $N_t=4$.}
 \label{random_WSR_Rth_b1}
 \end{figure}
Fig. \ref{random_WSR_Rth_b1} illustrates the WSR performance versus secrecy threshold  $R_{s,k}^{\rm th}$, when the transmit power constraint $P_t=20$dB, $N_t=2$ and $N_t=4$, respectively. As we can see, secure RS achieves an obvious WSR performance improvement compared with secure MULP in both sub-figures, since RS benefits from balancing between the common rate and the private rate. When $N_t=2$, as secrecy threshold $R_{s,k}^{\rm th}$ increases, the WSR of secure RS first remains constant and then decreases, whereas the WSR of secure MULP keeps decreasing at all time. That is because, adding secrecy constraints into the MULP scheme introduces a WSR performance loss comparing with that of no secrecy requirements as no common streams can be flexibly scheduled. However, when $N_t=4$, that is, the number of transmit antennas is larger than the total number of antennas at all users, although there is still an obvious performance difference between RS and MULP, both of them are able to ensure user security while keeping the WSR performance stable to some extent due to the reason that sufficient spatial resources are offered in such case.

\section{Conclusion}

In this paper, we investigate the secure beamforming design in RSMA-based multi-antenna broadcast channel. Taking user secrecy rate requirements and power constraint into consideration, a WSR maximization problem is formulated. To solve the non-convexity of the problem, a low-complexity SCA-based approach is adopted. Numerical results illustrate that, benefiting from the capability of RSMA for partially decoding MUI and partially treating MUI as noise, RSMA is able to maintain the WSR performance within a range of user security requirements. The WSR performance of RSMA  outperforms that of conventional MULP scheme while guaranteeing user secrecy rate requirements.

 %The impact of the adsorption rate and the desorption rate on the receiver performance is well characterized. 
\appendices

\bibliographystyle{IEEEtran}
\bibliography{sRS}
%\bibliography{IEEEabrv,ReceiverdoubleICCversion}

\balance
\end{document}